\documentstyle [12pt]{article}

\newcommand{\be}{\begin{equation}}
\newcommand{\ee}{\end{equation}}
\newcommand{\bea}{\begin{eqnarray}}
\newcommand{\ena}{\end{eqnarray}}

\def\(    {\left( }    \def\)   {\right) }
\def\[    {\left[}    \def\]   {\right] }
\def\<    {\big{\langle}}  \def\> {\big{\rangle}}
\def\half {\frac{1}{2}}
\def\txt #1 {\qquad {\mbox{#1}} \qquad}
\def\Tr      {\mbox{Tr}}
\def\Im     {\mbox{Im}}
\def\Re     {\mbox{Re}}
\def\lsim{\; \raise0.3ex\hbox{$<$\kern-0.75em\raise-1.1ex\hbox{$\sim$}}\; }
\def\gsim{\; \raise0.3ex\hbox{$>$\kern-0.75em\raise-1.1ex\hbox{$\sim$}}\; }
\def\@versim#1#2{\lower0.2ex\vbox{\baselineskip\z@skip\lineskip\z@skip
  \lineskiplimit\z@\ialign{$\m@th#1\hfil##\hfil$\crcr#2\crcr\sim\crcr}}}

\begin{document}
\begin{titlepage}
\rightline{CERN-TH.7336/94}

\bigskip
\bigskip
\bigskip
\bigskip

\begin{center}
{\bf {RESUMMATION OF PERTURBATION SERIES}}\\
{\bf {IN NON-EQUILIBRIUM SCALAR FIELD THEORY}}\\
\bigskip
\vspace{0.8cm}
{\large T. Altherr\footnote{On leave of absence from  L.A.P.P., BP110, F-74941
Annecy-le-Vieux Cedex, France.}}\\
{\em Theory Division, CERN, CH-1211 Geneva 23, Switzerland}\\
\medskip

{ \bf{Abstract}}\\
\end{center}
The general behaviour of perturbation series in non-equilibrium scalar field
theory is analysed in some detail, with a particular emphasis on the
``pathological terms'', generated by multiple products of
$\delta$-functions. Using an intuitive regularization method, it is shown that
these terms give large contributions at all orders, even when considering
small deviations from equilibrium. Fortunately, these terms can also be
resummed and I give the general expressions for the resummed propagators in
non-equilibrium scalar field theory, regardless of the size of deviations
from equilibrium.
 \bigskip

\centerline{\it Submitted to Physics Letters B}
\vfill

\leftline{CERN-TH.7336/94}
\leftline{June 94}
\end{titlepage}

\setcounter{footnote}{0}
\section{Introduction}
In a previous work \cite{AS}, Seibert and the present author showed
that the standard theoretical framework of non-equilibrium quantum field
theory is plagued with serious difficulties when perturbation
series is  considered. The problem originates from the
non-cancellation of the ``pathological terms'' \cite{LvW} that are associated
to multiple products of $\delta$-functions.  In \cite{AS}, it has been proved
that these terms do not cancel unless the particle distributions are those for
a
system in thermal and chemical equilibrium.

This paper is devoted to the discussion of a possible solution to this problem.
In particular, the behaviour of the perturbation series, that
is in terms of the coupling constant, is scrutinized in the light of the
presence
of these pathologies. First needed is a regularization
scheme,  in order to deal with the mathematically ill-defined pathologies.
Since the problem originates from the infinitesimal $i\epsilon$ prescription
in the free propagator, a rather natural way to perform this regularization is
to  introduce a finite width, or damping, in the  propagators.

Throughout the paper, the 2$\times$2 Keldysh matrix structure
\cite{NONEQ} is used as a  toy-model of a non-equilibrium quantum
field system, \bea
 &&\( \begin{array}{cc} D_{11}(K) & D_{12}(K) \\
                      D_{21}(K) & D_{22}(K) \end{array} \)
      = \nonumber\\
&&\(
           \begin{array}{cc}
\Delta_R(K)(1+n(k_0)) +\Delta_A(K)n(k_0) &
                                         (1+n(k_0))(\Delta_R(K)+\Delta_A(K))\\
 n(k_0)(\Delta_R(K)+\Delta_A(K))              &
 \Delta_R(K)n(k_0) +\Delta_A(K)(1+n(k_0))          \end{array} \)
,\label{PROP}\ena
with the Retarded/Advanced propagators defined as
\be
\Delta_{R(A)}(K) = {+(-) i\over K^2 - m^2 +(-) i\gamma k_0}
.\ee
The width $\gamma$ is an arbitrary finite function of $K$.
The distribution $n(k_0)$ is defined so that
\be
n(k_0) = -1-n(-k_0) = {\bf n} ({\bf k})
,\ee
where ${\bf n}({\bf k})$ represents the non-equilibrium distribution of  the
quanta,  namely
\be
\Tr \[ \rho\; a^{\dag}({\bf k}) a({\bf p}) \] = {\bf n}({\bf k}) \delta^3({\bf
k} - {\bf p}) ,\ee
where $\rho$ is an arbitrary density operator. It is assumed that the
system evolves sufficiently slowly in time and space so that the Fourier
transform, for short space-time separations, has the simple expression
shown in  eq.~(\ref{PROP}).

By introducing this finite width in the bare propagators, the
perturbation series now becomes mathematically well-defined:
pathologies associated with multiple products of $\delta$-functions are
regularized. In principle, this width, which is introduced heuristically ``by
hand'', can be calculated  by using perturbation theory. I shall discuss
this point in more detail later on.

The question is now: How do the pathological terms change the behaviour of
the perturbation series in terms of the coupling constant $g$?

\section{$g^2\phi^4$ theory}
{}From now on, I consider the massless case.
The simplest thing to compute in $(g^2/4!) \phi^4$ theory is the
tadpole diagram (in $n=4-2\epsilon$ dimensions)
\be
\Re\;\Sigma = {1\over 2} g^2 \int {d^n K\over (2\pi)^n} (1+2n(k_0))
                                    {\gamma k_0\over K^4 + \gamma^2 k_0^2}
.\ee
As usual, the ultraviolet singularities are present in the vacuum part only,
provided the distribution $n(k_0)$ drops sufficiently fast when $k_0 \to
\infty$, as in the equilibrium case. For the matter part (denoted by
$\delta\Sigma$ in the following), I obtain, after a Cauchy integration:
\be
\Re\;(\delta\Sigma) = {g^2 \over 8\pi^2} \int_0^\infty dk_0\;  k_0\;
n(k_0) \( \sqrt{1+i{\gamma\over k_0}} + \sqrt{1-i{\gamma\over k_0}} \)
,\label{1LOOP}\ee
where I have assumed, for the sake of simplification in the notation, that
$\gamma$ depends only on $k_0$. One recovers the well-known thermal
mass, $\delta m^2=(g^2/24) T^2$, when $\gamma\to 0$ and when
$n(k_0)$ is the Bose--Einstein distribution \cite{PHI4}. At one loop, the
imaginary part of $\Sigma$ vanishes, as in the case of zero width.

Hence, eq.~(\ref{1LOOP}) provides the complete non-equilibrium
generalization of the one loop thermal mass at equilibrium, regardless of the
particle number distribution $n(k_0)$. In the case of a small damping one has
\be
\delta m^2 =  {g^2 \over 4\pi^2} \int_0^\infty dk_0\;  k_0\;  n(k_0)
.\label{MASS}\ee
\bigskip

An inspection of the two loop diagrams does not reveal any serious infrared
problem. For the first topology, shown in fig.~1a, terms that are associated
with potential pathologies ($\delta^2(K^2) $ terms) cancel in the limit of
vanishing $\gamma$ \cite{AS}. Therefore, provided $n(k_0)$ is no more
singular than the Bose--Einstein distribution, the infrared problems are not
worst than at equilibrium. For instance, for a system that is close to
equilibrium, the two loop diagrams contribute at $O(g^3 T^2)$ to the
thermal mass \cite{PHI4}.

The same is true for the second topology, shown in fig.~1b, where no
pathologies are present.  This diagram has an imaginary part, which
also gives the width:
\be
\gamma(K) = {i\over 2k_0} \( \Sigma_{12}(K) - \Sigma_{21}(K)\)
.\ee
For small deviations from equilibrium, $\gamma$ is
just the usual equilibrium damping rate in $g^2 \phi^4$ theory, that is
\be
\gamma_{hard} = O(g^4 T)
,\label{GAMMAHARD}\ee
times eventually some $\ln{1/g}$ that arises from some logarithmic
infrared singularities (I do not discuss these minor corrections here).
The subscript $hard$ refers to the case of a hard external momentum
($K\sim T$). The soft ($K\sim gT$) damping rate is easily calculable and is
given by \be
\gamma_{soft} = {1\over 32\sqrt{6}\pi^3} g^3 \ln{1\over g} \; T
,\label{GAMMASOFT}\ee
to leading order in $\ln{1/g}$.
As motivated from the beginning, the dampings, either soft or hard,
eventually provide a way to regularize the pathological terms. However,
there remains to perform the resummation of the damping effects that lead to
the propagator (2) (which is only a useful assumption for the time being).
  \bigskip

Potential problems arise at the next order, that is at the three loop order,
which makes the analysis somewhat involved. As at the two loop order, the
repeated tadpole insertions do not lead to any pathological terms. So the
diagram shown in fig.~2a is no more singular than at equilibrium, and is of
$O(g^3 T^2)$, again for small deviations from equilibrium \cite{PHI4}. On
the other hand, pathological terms appear in the diagram shown in fig.~2b.
Upon using some useful relations between the different components of the
self-energy insertion $\Sigma^{1b}$ \cite{AS}, one may write
 \bea
-i\Sigma^{2b} &=&-i g^2 \int {d^n K\over (2\pi)^n}
\bigg\{ -i\Re\;\Sigma^{1b} (K)\[ D_{11}^2(K) - D_{12}(K)D_{21}(K) \]
\nonumber\\
&&-i\Sigma^{1b}_{12}(K)\[ -{1\over 2}(D_{11}^2(K) +
                                     D_{12}(K)D_{21}(K)) + D_{11}(K)D_{21}(K)\]
\nonumber\\
&&-i\Sigma^{1b}_{21}(K)\[ -{1\over 2}(D_{11}^2(K) +
                                     D_{12}(K)D_{21}(K)) + D_{11}(K)D_{12}(K)\]
\bigg\}.\ena
In terms of Retarded/Advanced propagators:
 \bea
\Sigma^{2b} &=& g^2 \int {d^n K\over (2\pi)^n}
\bigg\{ -i\Re\;\Sigma^{1b} (K)\[ (1+n(k_0))\Delta_R^2(K) -
n(k_0)\Delta_A^2(K) \]  \nonumber\\
&&+i\half\Sigma^{1b}_{12}(K)\[ (1+n(k_0))\Delta_R^2(K) +
n(k_0)\Delta_A^2(K) +2n(k_0)\Delta_R(K)\Delta_A(K)\] \nonumber\\
&&-i\half\Sigma^{1b}_{21}(K)\[ (1+n(k_0))\Delta_R^2(K) +
n(k_0)\Delta_A^2(K) +2(1+n(k_0))\Delta_R(K)\Delta_A(K)\]
\bigg\}.\label{3LOOP}\ena
The pathological terms are thus
 \bea
\Sigma^{2b} &=& g^2 \int {d^n K\over (2\pi)^n}
\Delta_R(K)\Delta_A(K) \[ -i\Sigma^{1b}_{12}(K)n(k_0)
                              +i\Sigma^{1b}_{21}(K) (1+n(k_0))\]
\nonumber\\
&&+\; \mbox{regular terms}
 .\ena
Let me now define a function $f$ such that
\be
 -i\Sigma^{1b}_{12}(K)n(k_0) +i \Sigma^{1b}_{21}(K) (1+n(k_0))
= g^4  f(k_0) \; \delta n(k_0)
,\label{ASUM}\ee
where $\delta n(k_0)=n(k_0)-n_B(k_0)$ represents the non-equilibrium
deviations. The pathological contribution to the three loop order reads
 \be
\Sigma^{2b} = {g^6\over 16\pi^2} \int_{-\infty}^{+\infty} dk_0
f(k_0) \; \delta n(k_0) {1\over \gamma}
\[ \sqrt{1+i{\gamma\over k_0}} + \sqrt{1-i{\gamma\over k_0}} \]
 ,\label{PATHO}\ee
which shows that, under the assumption that $f(k_0)$ and  $\delta n(k_0)$
do not have a singular infrared behaviour,
 \be
\Sigma^{2b} = O\( g^6 {\delta n(\kappa) \kappa^3 \over \gamma}\)
                     = O\( g^2 \delta n(\kappa) {\kappa^3 \over T}\)
,\ee
where $\kappa$ is the momentum scale that dominates the
integral in eq.~(\ref{PATHO}) (it is associated to the fluctuations). These
simple assumptions about the behaviour of the fluctuation terms
are just made to illustrate the possible breakdown of perturbation series in
the presence of the pathological terms. Let me further assume that the scale
$\kappa$ does not differ too much from the temperature $T$. Now, if
 \be \delta n(\kappa) = O(g^2) ,\ee
the pathological term is larger than the regular terms in
eq.~(\ref{3LOOP}) (that can be shown to be of $O(g^5)$). But it
is still smaller by $O(g)$ than the tadpole insertions diagram (fig.~3), which
is the dominant infrared diagram. This indicates that it might be possible to
get a control over the perturbation series in that case.

However, the above contribution is also of the same order, $O(g^4)$,  as the
non-equilibrium corrections to the first-order result (see eq.~(\ref{MASS})).
Therefore, nothing more than the equilibrium limit can be learned from
eq.~(\ref{MASS}) without calculating these terms.

Not surprisingly, it also turns out that the higher-order self-energy
insertions
contribute at the same order for the pathological term. After some algebra,
one can show that (see fig.~3):
 \bea
\Sigma^{(N)} &=& g^2 \int {d^n K\over (2\pi)^n}
\bigg\{   (1+n)\Delta_R\( -i\Sigma\Delta_R\) ^N
            + n\Delta_A \( +i\Sigma^*\Delta_A\) ^N
\nonumber\\
&& -i\Delta_R\Delta_A\( n\Sigma_{12} - (1+n)\Sigma_{21}\)
\sum_{n=0}^{N-1} \( -i\Sigma\Delta_R\) ^{N-1-n}
                                    \( +i\Sigma^*\Delta_A\) ^n
\bigg\}.\label{NLOOP}\ena
In order to simplify the notation, I have removed the $K$-dependence  in
all the quantities: $n(k_0)$, $\Delta(K)$ and $\Sigma(K)$. One important
remark must be added: the self-energy $\Sigma$ that enters in the above
expression must be a 2-point function, {\it it does not include the tadpole}.

 The usual way of dealing with this kind of formula is to use the
``mass-derivative formula'' \cite{MDF}, which  can be replaced by the
``width-derivative formula'' in the same fashion.  Equation (\ref{NLOOP})
shows one important feature: the pathological terms are always linear in
$\delta n$. A simple estimate leads to
\be \Sigma^{(N)} =O\( g^2\delta n(\kappa) \( T\over \kappa\) ^N
                                    {\kappa^4 \over T^2} \; \)
,\ee
where I have supposed that the $k_0$-integral is infrared safe (which is most
probably not the case, making the behaviour shown above rather
``optimistic"). Therefore, regardless of the respective size of the
pathological
terms versus the regular ones, it is clear that every repeated self-energy
insertions must be taken into account, as it can contribute to the same order
in the coupling constant.

The resummation of the non-pathological terms in
eq.~(\ref{NLOOP}) is a well-known exercise \cite{PHI4}. The nice surprise is
that the pathological terms can also be resummed, leading to
 \bea
\sum_{N=0}^\infty\Sigma^{(N)} &=& g^2 \int {d^n K\over (2\pi)^n}
\bigg\{   (1+n){\Delta_R\over 1+i\Sigma\Delta_R}
            + n{\Delta_A \over 1-i\Sigma^*\Delta_A}
\nonumber\\
&& -i\( n\Sigma_{12}-(1+n)\Sigma_{21}\)
{\Delta_R\over 1+i\Sigma\Delta_R} {\Delta_A\over 1-i\Sigma^*\Delta_A}
\bigg\}.\label{SUMLOOP}\ena
This is the central result. Besides the usual first terms that are similar to
the equilibrium resummation, the last term contains all the non-equilibrium
diseases that are cured in this rather elegant way. One can see that this term
behaves  as  eq.~(\ref{PATHO}),
\be
\Sigma_{pathology} = O(g^2\delta n)
.\ee
In principle, with this resummation, one is not restricted to the study
of systems that are close to equilibrium. For $\delta n > O(1)$, the
pathological term becomes the leading contribution to the mass shift, and
most probably to all other quantities.

Finally, I give the expression for the resummed propagator, which can be
read off from eq.~(\ref{SUMLOOP}):
\be
^*D_{11}(K)= (1+n(k_0))\; ^* \Delta_R(K) +
                                    n(k_0)\; ^*\Delta_A(K)
           + 2k_0\Delta n(k_0)\; ^* \Delta_R(K)\; ^*\Delta_A(K)
,\label{FULLPROP}\ee
where the $^*$ refers to the resummed quantities.
Note that  the following relation has been used \cite{Wel}
\bea
2k_0\Delta n(k_0) &=&
-i\( n(k_0)\Sigma_{12}(K)-(1+n(k_0))\Sigma_{21}(K)\)  \nonumber\\
&=&2k_0\( {dn(k_0)\over dt} + {\bf v}\cdot\nabla n(k_0) \)
,\ena
which is nothing but the quantum-field Boltzmann equation \cite{NONEQ}.
For completeness, I also list the resummed expression for the other matrix
components \bea
^*D_{22}(K) &=& (\; ^*D_{11}(K))^*  \nonumber\\
^*D_{12}(K) &=& (1+n(k_0)) (\; ^* \Delta_R(K) +\; ^*\Delta_A(K))
           + 2k_0\Delta n(k_0)\; ^* \Delta_R(K)\; ^*\Delta_A(K)
\nonumber\\
^*D_{21}(K) &=& n(k_0) ( \; ^* \Delta_R(K) + \; ^*\Delta_A(K))
           + 2k_0\Delta n(k_0)\; ^* \Delta_R(K)\; ^*\Delta_A(K)
.\ena
 \section{$g \phi^3$ theory}
For completeness, I also analyse the behaviour of perturbation series in
$(g/3!)\phi^3$ theory (in 6 space-time dimensions).  It is evident that the
situation can  be different, as the repeated self-energy insertions on the same
line  contain some pathological terms and is at the same time the leading
infrared series in the equilibrium limit.

Let me begin with the real part of the one loop self-energy. As the interest
here is to make  a possible generalization to the case of gauge theories, I
discard the tadpole diagram from the discussion \cite{PHI3}. One has,
 \be
\Re\;\Sigma (P)= {1\over 2} g^2 \int {d^n K\over (2\pi)^n}
                          2 \Re\; D_{11}(K) \; \Im\; D_{11}(P-K)
.\ee
In order to simplify the calculations, I use the ``Hard Thermal Loop''
approximation, that is I consider small external momenta and I suppose
that the dominant contribution to the integral is such that $P\ll K$.
Using the same techniques as in the $g^2\phi^4$ case, I easily obtain
\be
 \Re\;(\delta\Sigma(P)) = -{g^2 \over 64\pi^3} \int_0^\infty dk_0\;
k_0\; n(k_0) \( \sqrt{1+i{\gamma\over k_0}} + \sqrt{1-i{\gamma\over
k_0}} \) .\label{PHI3MASS}\ee
Again, one recovers the well-known thermal mass,
$\delta m^2=-g^2/(384\pi) T^2$, when $\gamma\to 0$ and when
$n(k_0)$ is the Bose--Einstein distribution \cite{PHI3}.

Now, pathological terms arise already at the two loop level. This is easily
seen
in the diagram shown in Fig.~4, which gives
 \bea
\Sigma^{4}_{11}(P) &=&-ig^4 \int {d^n K\over (2\pi)^n}
\Delta_R(K)\Delta_A(K) \nonumber\\
&&\times\[ -i\Sigma^{(1)}_{12}(K)n(k_0)
                              +i\Sigma^{(1)}_{21}(K) (1+n(k_0))\] D_{11}(P-K)
\nonumber\\
&&+ \; \mbox{regular terms}
 .\ena
Using the same assumptions as in the previous case (eq.~(\ref{ASUM})), I
obtain (in HTL approximation):
 \be
\Re\;\Sigma^{4}_{11} = -{g^4\over 64\pi^3} \int_{-\infty}^{+\infty}
dk_0 f(k_0) \; \delta n(k_0) {1\over \gamma}
\[ \sqrt{1+i{\gamma\over k_0}} + \sqrt{1-i{\gamma\over k_0}} \]
 .\ee
For small deviations from equilibrium, the damping is of order $g^4 T$,
and therefore
\be
\Re\;\Sigma^{4}_{11} = O(\delta n)
.\ee
Although the breakdown of perturbation series seems worse than in the
previous case, this is not so. The reason for this is the regularization scheme
that is not adequate here. Indeed, the resummed propagator that includes all
the pathological terms is the same as in the previous case (see
eq.~(\ref{FULLPROP})). But there is now a non-vanishing contribution at
$O(g^2)$ for the 2-point Green's function (this is just the thermal mass,
eq.~(\ref{PHI3MASS})), which acts as an infrared regulator. Therefore the
above contribution is much smaller \be
\Re\; \Sigma^{4}_{11} = O(g^2\delta n)
,\ee
that is the same as in the $g^2\phi^4$ case.
Hence, I end up with the same conclusions. Pathological terms will be the
leading effects whenever considering deviations from equilibrium such that
$\delta n > O(1)$.

\section{Conclusion and outlook}
The behaviour of perturbation series in the presence of pathologies,
associated to multiple products of $\delta$-functions,  has been analysed in
some detail . Although the regularization method used here is very heuristic, I
believe that it illustrates well the fact that, even for small deviations from
equilibrium, the perturbation series break down. Anyway, the central result
of this work is to show that these large contributions can be resummed. The
resulting theory is free of pathologies and set up a coherent framework for
the study of non-equilibrium systems, even in the case of large deviations
from equilibrium.

Of course, depending on the distribution $n(k_0)$, some additionnal
infrared problems may show up, but this needs a case-by-case study.

As usual in thermal field theory, the results that  have been found here in
scalar theory should easily translate to the case of gauge theories. There
remains to be seen whether or not gauge-invariance problems show up when
using a resummed propagator such as eq.~(\ref{FULLPROP}).

\section*{Acknowledgements}
I wish to thank F.~Gelis, R.~Kobes, T.~del~Rio~Gaztelurrutia and
D.~Seibert for discussions.

\newpage
{\large\bf Figure captions\\[0.5cm]}
\begin{description}
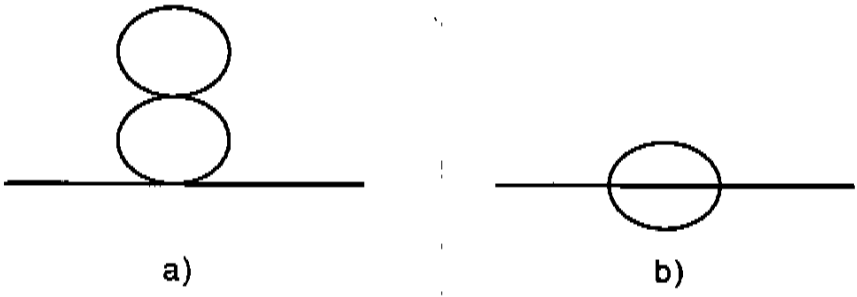
\item[Fig.~1] The two loop order contribution to the self-energy in
$g^2\phi^4$.
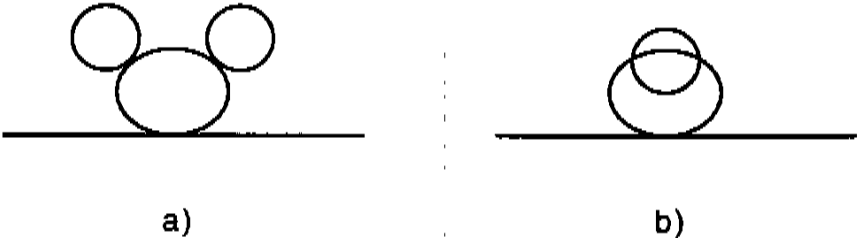
\item[Fig.~2] The leading infrared three loop order contribution to the
self-energy in $g^2\phi^4$.
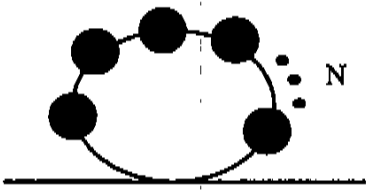
\item[Fig.~3] $N$ self-energy insertions on the tadpole loop.
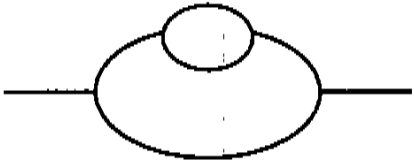
\item[Fig.~4] A two loop order contribution to the self-energy in
$g\phi^3$.
\end{description}

 \end{document}